\documentclass[aip, jcp, reprint]{revtex4-1}
\usepackage{graphicx}
\usepackage{amsmath}
\usepackage{float}
\begin{document}
 \floatstyle{plain}
 \newfloat{scheme}{thp}{los}
 \floatname{scheme}{\small{SCHEME} }
\title{Quasi-classical trajectories study of Ne$_2$Br$_2$($B$) vibrational
predissociation: Kinetics and product distributions}

\author{Wilmer Arbelo-Gonz\'alez}
\affiliation{Departamento de F\'{\i}sica General, Instituto Superior de
Tecnolog\'{\i}as y Ciencias Aplicadas, Habana 10600, Cuba}

\author{Maykel L.~Gonz\'alez-Mart\'{\i}nez}
\email{m.l.gonzalez-martinez@durham.ac.uk}
\altaffiliation{Current address: Department of Chemistry, Durham University,
Durham DH1~3LE, United Kingdom}
\affiliation{Departamento de F\'{\i}sica General, Instituto Superior de
Tecnolog\'{\i}as y Ciencias Aplicadas, Habana 10600, Cuba}

\author{Stewart K.~Reed}
\affiliation{School of Chemistry, University of Leeds, Leeds LS2~9JT, United
Kingdom}

\author{Jes\'us Rubayo-Soneira}
\affiliation{Departamento de F\'{\i}sica General, Instituto Superior de
Tecnolog\'{\i}as y Ciencias Aplicadas, Habana 10600, Cuba}

\author{Dmitrii V.~Shalashilin}
\affiliation{School of Chemistry, University of Leeds, Leeds LS2~9JT, United
Kingdom}

\begin{abstract}
The vibrational predissociation of the Ne$_2$Br$_2$(\emph{B}) van der Waals
complex has been investigated using the quasi-classical trajectory method (QCT),
in the range of vibrational levels $v' = 16$--23.  Extensive comparison is made
with the most recent experimental observations [Pio \emph{et al.}, J.\ Chem.\
Phys.\ \textbf{133}, 014305 (2010)], molecular dynamics with quantum transitions
(MDQT) simulations [Miguel \emph{et al.}, Faraday Discuss.\ \textbf{118}, 257
(2001)], and preliminary results from 24-dimensional Cartesian coupled coherent
state (CCCS) calculations.  A sequential mechanism is found to accurately
describe the theoretical dynamical evolution of intermediate and final product
populations, and both QCT and CCCS provide very good estimates for the
dissociation lifetimes.  The capabilities of QCT in the description of the
fragmentation kinetics is analyzed in detail by using reduced-dimensionality
models of the complexes and concepts from phase-space transport theory.  The
problem of fast decoupling of the different coherent states in CCCS simulations,
resulting from the high dimensionality of phase space, is tackled using a
re-expansion scheme.  QCT ro-vibrational product state distributions are
reported.  Due to the weakness of the vdW couplings and the low density of
vibrational states, QCT predicts a larger than observed propensity for $\Delta
v' = -1$ and $-2$ channels for the respective dissociation of the first and
second Ne atoms.
\end{abstract}



\maketitle

\section{Introduction}
\label{sec:introduction}
Studying the influence of size-selected solvents on the structure and dynamics
of molecular systems is essential in understanding both molecular energy
transfers and the transition from the gas to condensed phase.

In this regard, clusters of rare gas (Rg) atoms doped with a diatomic halogen
(BC) are particularly convenient for at least two reasons: (1) the weakness of
the van der Waals (vdW) interactions provides a means to effectively `separate'
the diatom from the environment, which in turn allows a relatively simple
identification of the different energy transfer mechanisms at the state-to-state
level; and (2) the number $n$ of Rg atoms can be spectroscopically selected for
their addition induces a known blueshift in the vibronic transition
$B(v') \leftarrow X(v=0)$ \cite{wsharfin:79,jekenny:80,baswartz:84}.

These weakly bound complexes have been the subject of intense scrutiny since the
pioneering experiments of Levy and co-workers \cite{resmalley:76,gkubiak:78,
wsharfin:79,jekenny:80,kejohnson:81} and the theoretical work of Beswick,
Jortner and Delgado-Barrio \cite{jabeswick:79a,jabeswick:79b} in the late
seventies.  In their experiments, Levy and co-workers used laser-induced
fluorescence to study Rg$_n$I$_2$ molecules based on He, Ne and Ar with up
$n = 7$ I atoms.

Following the laser-induced vibrational excitation of BC, the energy is usually
redistributed within the molecule leading to the breaking of the vdW bonds.
This process is known as vibrational predissociation (VP) and provides
significant information, for instance, on the dynamics of intramolecular
vibrational energy redistribution (IVR).  Rotational, electronic and other
predissociation processes are possible.  In the former, rotational de-excitation
of the BC molecule directly provides enough energy for fragmentation.  In
electronic predissociation, non-adiabatic transitions to repulsive electronic
states of BC may instead lead to dissociation of the chemically bonded molecule.
This was first observed in Ar$_n$I$_2$ \cite{gkubiak:78,kejohnson:81} and
recently studied in HeBr$_2$ and NeBr$_2$ \cite{mataylor:10}.  In general, the
dependence of the transitions on the details of the potential energy surfaces
(PESs) as well as on the available energy makes it possible to extract useful
information on binding energies and electronic couplings.

The VP of vdW complexes is by far the predominant dissociation process and as
such has been the most extensively studied.  Depending on the vibrational state
excited, vdW systems fragment following one or several elementary steps, which
usually include direct dissociation, different IVR regimes, evaporative cooling
(EC), etc.  For example, the VP of Ne$_n$Br$_2$ clusters is described in
references \onlinecite{bmiguel:00,bmiguel:01}.  Each step may have different
accessible final states and characteristic kinetics which manifest themselves in
experiments through distinctively structured product state distributions as in
the case of ArCl$_2$ \cite{ddevard:88}, HeBr$_2$ and NeBr$_2$
\cite{arohrbacher:99b,mnejad-sattari:97}.

All of the above work (and considerably more) has proved through the years that,
despite their apparent simplicity, even small vdW aggregates with $n = 1, 2$
undergo a wide variety of processes that are representative of most of the
dynamical pathways observed in more complex, and conventionally bonded,
molecules \cite{arohrbacher:00}.  Hence, vdW systems have become valuable
prototypes in the analysis of both inter- and intramolecular energy transfers.

For small molecules, exact quantum mechanical calculations (EQM) can be
performed with modern computers and algorithms.  Being essentially exact, at
least to the accuracy of the PESs, these calculations have provided a rigorous
picture of the VP process.  More recently, they have also addressed in detail
the problem of IVR dynamics and the role played by continuum resonances in
triatomic systems \cite{tglez-lezana:96,tglez-lezana:97,tastephenson:00,
oroncero:01,agarcia-vela:06,agarcia-vela:07,agarcia-vela:08}.  Quantum
calculations on larger complexes almost inevitably use various approximate
methods and there are just a few EQM studies in the literature.  In 2001, Meier
and Manthe studied the VP of Ne$_2$I$_2$ using the multiconfiguration
time-dependent Hartree method\cite{cmeier:01} and although vibrational branching
ratios were compared with the experiment \cite{jekenny:80}, the main aim of
their work was to provide a benchmark for future developments in the
methodology.  In 2005, Garc\'{\i}a-Vela proposed a full-dimensional,
fully-coupled wave packet method and used it to study the VP of He$_2$Cl$_2$
\cite{agarcia-vela:05}, obtaining good agreement with experimental lifetimes and
rotational distributions thus providing a test for the accuracy of the PES
employed.

Yet, time-dependent, fully detailed investigations of energy transfers in vdW
molecules containing more than one or two Rg atoms still pose a considerable
challenge for both experiments and EQM calculations.  In the former, researchers
should be able to identify and characterize all intermediate complexes, as well
as address their dynamical evolution.  In the latter they have to cope with the
increasing number of degrees of freedom (DOF) and large basis sets that
eventually make the computational cost prohibitive.  Thus, in the investigation
of larger systems hybrid quantum-classical, \emph{e.g.}~molecular dynamics with
quantum transitions (MDQT) simulations \cite{abastida:98,sfdez-alberti:99,
abastida:99b,bmiguel:00,bmiguel:01}, and quasi-classical trajectory (QCT)
methods have been to date the most widely used practical alternatives.

In fact, most QCT applications have successfully reproduced the fragmentation
kinetics for at least several tri- and tetra-atomic complexes.  This is the case
for molecules such as Ne$_n$I$_2$ ($n=$~1, 2) \cite{jrubayo-soneira:95a,
agarcia-vela:96,oroncero:01}, HeICl \cite{rlwaterland:88} and NeBr$_2$
\cite{mnejad-sattari:97,mlglez-mtnez:06a} in which many classical predictions
were later confirmed by EQM \cite{rlwaterland:90}.  More recently, some of us
have explored the extent to which the Gaussian weighted (GW) trajectory method
\cite{lbonnet:97,lbonnet:04a} adds to the applicability of QCT in the study of
RgBr$_2$ (Rg = He, Ne, Ar) molecules \cite{mlglez-mtnez:08}.  We found that GW
may enhance the QCT description of product state distributions both quantitative
and qualitatively, especially if only a few vibrational states are populated or
calculations are performed very close to a channel closing.  However, in cases
such as the VP of NeCl$_2$ \cite{jicline:89} and generally, when IVR takes place
in the sparse regime and plays a significant role, it seems that only quantum
mechanical calculations can be compatible with the experimental observations.
Hence, despite the success of many previous implementations of QCT, the validity
of classical concepts in the context of the VP of vdW aggregates is yet to be
clarified.  At last, complementary studies addressing the dynamical
(in)stability of these systems exist in the literature.  The analysis of
phase-space bottlenecks in the predissociation of HeI$_2$ was used as a
benchmark in the foundation of phase-space transport theory (PSTT)
\cite{mjdavis:86,skgray:87,regillilan:91,mzhao:92a,mzhao:92b}.  More recently,
irregular variations of decay rates as well as details of the absorption spectra
have been studied through the analysis of the phase-space structure and its
evolution with increasing excitation energy \cite{aagranovsky:98,rprosmiti:02c,
rsospedra-alfonso:03,mlglez-mtnez:05a}.

In general, testing the accuracy and reliability of theoretical methods largely
depends on detailed experimental data becoming available.  In particular, as
stated above, realistic kinetic mechanisms can only be obtained if the dynamics
of all intermediate species is recorded in the experiment. This has
traditionally proved to be quite challenging.  For instance, in 1992, Gutmann
\emph{et al.}\ \cite{mgutmann:92} used picosecond pump-probe spectroscopy to
study the VP of Ne$_n$I$_2$ with $n=$~2--4 but only registered the formation of
I$_2$.  They therefore inferred the evolution of intermediates by fitting the
product dihalogen formation to a sequential first-order kinetic mechanism.  In
2010, Pio \emph{et al.}\ \cite{jmpio:10} reported the characterization of the VP
of Ne$_2$Br$_2$($B$) at an unprecedented level of detail.  Using time- and
frequency-resolved pump-probe spectroscopy they were able to record the real
time evolution of all complexes involved, propose a kinetic mechanism, and
determine time constants and product vibrational state distributions.

In this paper, we report QCT and preliminary Cartesian coupled coherent states
(CCCS) \cite{skreed:10a,skreed:11a} calculations on the VP of
Ne$_2$Br$_2$($B$, $v' =$~16--23).  We compare both theoretical methods, as well
as their ability to reproduce the experimental observations \cite{jmpio:10} and
previous MDQT predictions \cite{bmiguel:01}.  Our main goal is to build upon the
current understanding of the capabilities of QCT in the simulation of VP
processes involving larger vdW clusters, and to distinguish whenever possible
the results which are intrinsic to the methodology from those that are
characteristic to the systems under consideration.  Complementary analysis of
full-dimensional and simplified models for the VP of the NeBr$_2$ triatomic
complex have been very useful in the interpretation of QCT results and are
conveniently discussed.

\section{Theory}
\label{sec:theory}
\subsection{Vibrational predissociation of a tetra-atomic van der Waals cluster}
\label{sec2:VP_Rg1Rg2BC}

\begin{scheme*}
 \begin{eqnarray*}
  \mathrm{Rg}_1 \mathrm{Rg}_2 \mathrm{BC}(X,v,j)
                         &\stackrel{h\nu}{\longrightarrow}&
  \left[ \mathrm{Rg}_1 \mathrm{Rg}_2 \mathrm{BC}(B,v',j') \right]^* \\
                         &\stackrel{\mathrm{VP}_{(1)}}{\longrightarrow}&
  \left\{ \begin{array}{l}
   (S_{1(2)})\; \mathrm{Rg}_{1(2)} +
       \mathrm{Rg}_{2(1)} \mathrm{BC}(B,v^\mathrm{i},j^\mathrm{i})
          \stackrel{\mathrm{VP}_{(2)}}{\longrightarrow}
           \mathrm{Rg}_{1} + \mathrm{Rg}_{2}
           + \mathrm{BC}(B,v^\mathrm{f},j^\mathrm{f}) \\
   (C_1)\; \mathrm{Rg}_{1} + \mathrm{Rg}_{2}
           + \mathrm{BC}(B,v^\mathrm{f},j^\mathrm{f}) \\
   (C_2)\; \mathrm{Rg}_{1}\mathrm{Rg}_{2}(v^\mathrm{f}_{12},j^\mathrm{f}_{12})
           + \mathrm{BC}(B,v^\mathrm{f},j^\mathrm{f})
          \end{array} \right.
 \end{eqnarray*}
 \caption{\raggedright Representation of the VP of a generic Rg$_1$Rg$_2$BC vdW
  molecule, showing the ($S_{1(2)}$) sequential; as well as concerted ($C_1$)
  \emph{without}; and ($C_2$) \emph{with} molecular formation mechanisms.
  \label{sch:VP_Rg1Rg2BC}}
\end{scheme*}

Assuming that the photo-excitation does not provide enough energy to break the
BC bond, the VP of a tetra-atomic vdW aggregate is represented in
Scheme~\ref{sch:VP_Rg1Rg2BC} where the i and f superscripts denote intermediate
and final states respectively.  Two main fragmentation paths are possible, which
we refer to as the \emph{sequential} ($S$) and \emph{concerted} ($C$)
mechanisms, comprising two channels each.  The former is associated with two
well-defined dissociation steps and usually dominates as long as direct
dissociation prevails over IVR.  It leads to the loss of two vibrational quanta
and have relatively simple implications in the VP kinetics \cite{bmiguel:01,
agarcia-vela:96}.  The latter becomes increasingly important as IVR dominates,
and is very often linked to the loss of more than two vibrational quanta and
highly-structured product state distributions.  All these features help in
distinguishing between the concerted and sequential mechanisms for in the vast
majority of cases they lead to the same final products (compare $S_{1(2)}$ and
$C_1$).  Finally, when the interaction between the Rg atoms is sufficiently
strong and the dynamical evolution allows for favorable configurations, there is
a non-negligible probability of formation of a Rg$_1$Rg$_2$ molecule.

\subsection{Potential energy surfaces}
\label{sec2:pess}
As seen in Scheme~\ref{sch:VP_Rg1Rg2BC} both the ground $X$ and excited $B$
electronic states are, at least in principle, involved in the VP process.
Recent studies on the structure of He$_2$Br$_2$ and HeICl complexes
\cite{avaldes:05,avaldes:06} and EQM calculations on He$_2$Cl$_2$
\cite{agarcia-vela:05} have shown that the global PES for these systems is
accurately approximated by
\begin{eqnarray}
 V = V_\mathrm{Rg_1,BC} + V_\mathrm{Rg_2,BC} + V_\mathrm{Rg_1 Rg_2} +
  V_\mathrm{BC},
 \label{eq:global_pes}
\end{eqnarray}
where $V_\mathrm{BC}$ is the interaction potential of the isolated BC molecule,
$V_\mathrm{Rg_i,BC}$ ($i=1,2$) the vdW PES of the $i$th triatomic aggregate and
$V_\mathrm{Rg_1 Rg_2}$ the potential describing the Rg$_1$--Rg$_2$ interaction.

\begin{table}[!t]
 \begin{center}
  \caption{Morse parameters for the various pair interactions in Ne$_2$Br$_2$.
   \label{tab:Morse}}
  \begin{tabular}{ccrrrc}
   \hline\hline
          & Br$_2$ state & $D$/cm$^{-1}$ & $\alpha$/\AA$^{-1}$ & $r_\mathrm{eq}$/\AA & Ref.                          \\ \hline
   Br--Br & $X$          & 24,557.674    & 1.588               & 2.281               & \onlinecite{tglez-lezana:96}  \\
          & $B$          &  3,788.0      & 2.045               & 2.667               & \onlinecite{rfbarrow:74}      \\ \hline
   Br--Ne & $X$          &     45.0      & 1.67                & 3.7                 & \onlinecite{agarcia-vela:06}  \\
          & $B$          &     42.0      & 1.67                & 3.9                 & \onlinecite{aabuchachenko:94} \\ \hline
   Ne--Ne & $X$, $B$     &     29.36     & 2.088               & 3.091               & \onlinecite{agarcia-vela:96}  \\
   \hline\hline
  \end{tabular}
 \end{center}
\end{table}

Here, we extrapolate these results and in a first step, we express the global
PESs for the Ne$_2$Br$_2$ complex in the form \eqref{eq:global_pes}.  It is
important to note that, even if this proves to be a good approximation for Rg =
He and Ne due to the weakness of the vdW bonds, neglecting 4-body contributions
in the global PES of clusters containing heavier Rg atoms could eventually fail.
Secondly, by using pairwise additive potentials for the terms
$V_\mathrm{Rg_i,BC}$, we neglect 3-body contributions to the triatomic vdW PES.
Although this is known to be a good approximation for the $B$ electronic state,
which is where the VP process takes place, it is only partially adequate for the
ground state \cite{fynaumkin:95,arohrbacher:99a}.  In particular, there is
theoretical \cite{rprosmiti:02b} and experimental \cite{jmpio:08} evidence for
an additional minimum at linear configurations in the PES of the $X$ state,
which cannot be reproduced by pairwise interactions.  The topology of the two
electronic states is in fact similar but the linear minimum for the $B$ state
moves to a longer distance and becomes much shallower compared to the $X$ state,
which makes a pairwise additive description adequate for the former and not the
latter.  Nevertheless, the calculations presented here refer to the
fragmentation induced by the photo-excitation of the T-shaped isomer which can
be correctly reproduced in the pairwise additive approximation.  All pair
interactions are analytically modeled by Morse functions, the parameters for
which have been taken from the literature and are summarized in
Table~\ref{tab:Morse}.

\subsection{Quasi-classical trajectories}
\label{sec2:qct}
\subsubsection{System model}
\label{sec3:qct_system}
A particular Jacobi association diagram yields the most convenient set of
coordinates to describe the unimolecular dissociation of a triatomic RgBC vdW
complex.  Two Jacobi vectors are involved, $\boldsymbol{r}$ which conventionally
runs from the heaviest to the lightest of atoms B or C, and $\boldsymbol{R}$,
from the diatom's center of mass towards the Rg atom.  Among its main advantages
are the symmetrical decomposition of the molecular PES, the explicit use of a
dissociation coordinate and an associated diagonal kinetic operator.  It is
therefore common when studying larger vdW aggregates,
\emph{i.e.}~Rg$_1 \ldots$Rg$_n$BC, to choose a `generalized' set of vectors
$\boldsymbol{r}$, $\boldsymbol{R}_i$ ($i=\overline{1,n}$) which inherits most of
the aforementioned advantages.  These are not actually Jacobi co-ordinates but
are referred to as satellite or bond co-ordinates.  Their main drawback being a
non-diagonal kinetic operator.

\begin{figure}[!t]
 \begin{center}
  \includegraphics[width=85mm]{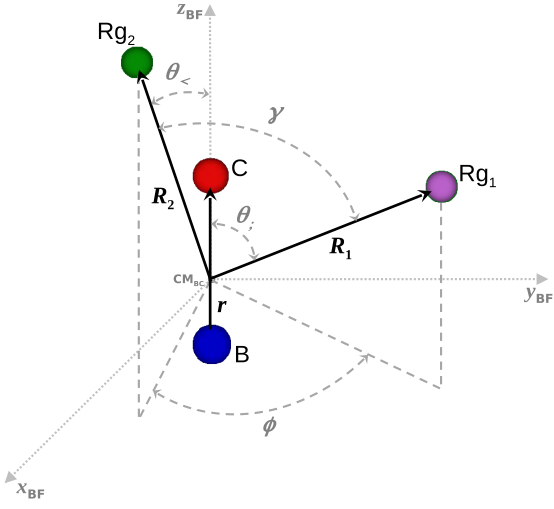}
  \caption{(Color online) Coordinate set for a prototypical Rg$_1$Rg$_2$BC vdW
   molecule in its body-fixed reference frame.
   \label{diag:qct_coordinates}}
 \end{center}
\end{figure}

The total angular momentum of a tetra-atomic vdW system is given by
$\boldsymbol{J} = \boldsymbol{l}_1 + \boldsymbol{l}_2 + \boldsymbol{j}$, where
$\boldsymbol{l}_i$ is the end-over-end orbital angular momentum of atom Rg$_i$
with respect to BC, and $\boldsymbol{j}$ the rotational angular momentum of BC.
Using satellite coordinates and keeping $\boldsymbol{J}$ at zero, the set of
coordinates reduces to $\{r,R_1,R_2,\theta_1,\theta_2,\phi\}$.  These have been
depicted in Fig.~\ref{diag:qct_coordinates}.  Choosing  $\boldsymbol{J}$ to be
zero is a well-justified constraint when studying the photodissociation of
rotationally cold species \cite{rschinke:book93} as produced in the experiment
through a supersonic free jet expansion \cite{jmpio:10}.  The complete set of
variables in classical phase space $\mathbf{\Gamma}$ is finally obtained by
adjusting the respective conjugate momenta so that the Hamilton function reads
\cite{jrubayo-soneira:phd}
\begin{eqnarray}
 \mathcal{H}^{\boldsymbol{J}=\boldsymbol{0}}&=&\mathcal{H}_\mathrm{BC}+
                                   \sum^{2}_{i=1}\frac{1}{2\mu_\mathrm{Rg_i,BC}}
                        \left(P^2_{R_i}+\frac{\boldsymbol{l}^2_i}{R^2_i}\right)+
                                   \frac{\boldsymbol{P}_1\cdotp\boldsymbol{P}_2}
                                         {m_\mathrm{B}+m_\mathrm{C}} \nonumber\\
            &&+V_\mathrm{vdW}(\boldsymbol{r},\boldsymbol{R}_1,\boldsymbol{R}_2),
\label{eq:qct_hamiltonian}
\end{eqnarray}
where
\begin{equation}
 \mathcal{H}_\mathrm{BC}\stackrel{\mathrm{def}}{=}
      \frac{1}{2\mu_\mathrm{BC}}\left(P^2_r+\frac{\boldsymbol{j}^2}{r^2}\right)+
                                                                V_\mathrm{BC}(r)
\label{eq:qct_HBC}
\end{equation}
and
\begin{equation}
 V_\mathrm{vdW}(\boldsymbol{r},\boldsymbol{R}_1,\boldsymbol{R}_2)
  \stackrel{\mathrm{def}}{=}V(\boldsymbol{r},\boldsymbol{R}_1,\boldsymbol{R}_2)-
                                                               V_\mathrm{BC}(r).
\label{eq:qct_VvdW}
\end{equation}
Here,
$\mu_\mathrm{Rg_i,BC}^{-1}=m_\mathrm{Rg_i}^{-1}+(m_\mathrm{B}+m_\mathrm{C})^{-1}
$ and $\mu_\mathrm{BC}^{-1}=m_\mathrm{B}^{-1}+m_\mathrm{C}^{-1}$ are the inverse
of the appropriate reduced masses, while the angular momenta
\begin{equation}
 \boldsymbol{l}^2_i=P^2_{\theta_i}+\frac{P^2_\phi}{\sin^2{\theta_i}};\quad i=1,2
\label{eq:qct_li}
\end{equation}
\begin{widetext}
and
\begin{eqnarray}
 \boldsymbol{j}^2&=&\boldsymbol{l}^2_1+\boldsymbol{l}^2_2+
                             2\boldsymbol{l}_1\cdotp\boldsymbol{l}_2
  = P^2_{\theta_1}+P^2_{\theta_2}+2P_{\theta_1}P_{\theta_2}\cos{\phi}
          -2\sin{\phi}\left(\frac{\cos{\theta_2}}{\sin{\theta_2}}P_{\theta_1}+
      \frac{\cos{\theta_1}}{\sin{\theta_1}}P_{\theta_2}\right)P_\phi \nonumber\\
                 &&+\left(\frac{1}{\sin^2{\theta_1}}+\frac{1}{\sin^2{\theta_2}}-
  2\cos{\phi}\frac{\cos{\theta_1}\cos{\theta_2}}{\sin{\theta_1}\sin{\theta_2}}-
                                                               2\right)P^2_\phi.
\label{eq:qct_j}
\end{eqnarray}
The non-diagonal coupling term, $\boldsymbol{P}_1\cdotp\boldsymbol{P}_2$, can
be expressed as
\begin{eqnarray}
 \boldsymbol{P}_1\cdotp\boldsymbol{P}_2&=&P_{R_1}P_{R_2}\cos{\gamma}-
     \frac{\cos{\phi}}{R_1R_2\sin{\theta_1}\sin{\theta_2}}P^2_{\phi} \nonumber\\
   &&+\frac{\cos{\theta_1}\cos{\theta_2}\cos{\phi}+\sin{\theta_1}\sin{\theta_2}}
                                               {R_1R_2}P_{\theta_1}P_{\theta_2}+
      \frac{\sin{\theta_1}\cos{\theta_2}\cos{\phi}-\cos{\theta_1}\sin{\theta_2}}
                                            {R_2}P_{R_1}P_{\theta_2} \nonumber\\
   &&+\frac{\cos{\theta_1}\sin{\theta_2}\cos{\phi}-\sin{\theta_1}\cos{\theta_2}}
                                                       {R_1}P_{R_2}P_{\theta_1}-
              \frac{\sin{\theta_1}\sin{\phi}}{R_2\sin{\theta_2}}P_{R_1}P_{\phi}-
   \frac{\sin{\theta_2}\sin{\phi}}{R_1\sin{\theta_1}}P_{R_2}P_{\phi} \nonumber\\
   &&-\frac{\cos{\theta_1}\sin{\phi}}{R_1R_2\sin{\theta_2}}P_{\theta_1}P_{\phi}-
      \frac{\cos{\theta_2}\sin{\phi}}{R_1R_2\sin{\theta_1}}P_{\theta_2}P_{\phi},
\label{eq:qct_P1P2}
\end{eqnarray}
\end{widetext}
with
\begin{equation}
 \cos{\gamma}=\sin{\theta_1}\sin{\theta_2}\cos{\phi}+
                                                   \cos{\theta_1}\cos{\theta_2}.
\label{eq:qct_cosgamma}
\end{equation}

Finally, in the interpretation of our results, models with two and three DOFs
(2/3-DOF) are used for the intermediate RgBC complex.  The 3-DOF model is a
full-dimensional $\boldsymbol{J} = \boldsymbol{0}$ approximation of RgBC.  The
phase-space variables span a subspace of $\mathbf{\Gamma}$,
\emph{i.e.}~$\mathbf{\Gamma}_\mathrm{3} = \{r, R, \theta, P_r, P_R, P_\theta\}$,
and all relevant formulae can be easily obtained from
Eqs.~\eqref{eq:qct_hamiltonian}--\eqref{eq:qct_VvdW} or the literature,
\emph{e.g.}~Ref.~\onlinecite{mlglez-mtnez:08}.  The 2-DOF is constructed by
additionally fixing $\theta = \pi/2$,  corresponding to the equilibrium
configuration of NeBr$_2$ in the $B$ electronic state.  The phase state is thus
$\mathrm{\Gamma}_\mathrm{2} = \{r, R, P_r, P_R\}$ and all necessary formulae
can be obtained by simplification of the 3-DOF.  In particular, several results
can be better understood by analyzing the structure of classical phase space for
the 2-DOF model.  To this end, we have used Poincar\'e surfaces of section
(SOS), which are a powerful visual tool when considering systems with two DOF
\cite{ajlichtenberg:book93}.  All SOS employed here were constructed from
trajectory intersections with the hypersurface $(r = r_\mathrm{eq}, R, P_r \ge
0, P_R)$, where $r_\mathrm{eq}$ is the equilibrium bond length of BC.

\subsubsection{Initial Conditions}
\label{sec3:qct_ic.theory}
In general, the QCT simulation of photo-induced processes requires the initial
conditions to closely match those recreated in the experiment
\cite{rschinke:book93}.  The latter are however quantum in nature, and are often
quite difficult to determine and reproduce classically, especially
for polyatomic systems.  Obviously, the workarounds commonly used in systems
with strong inter-/intramolecular interactions\footnote{After assuming that
strong IVR makes all available configurations equally populated within an
induction time relatively short compared to the time scale of the process of
interest, classical initial conditions are computed using microcanonical
distributions.} are not applicable here and quantum distributions need to be
calculated from the molecular wave function.

Although the rigorous form of the initial wave function is known to depend on
the particular shape of the laser pulse,
\emph{cf.}~Ref.~\onlinecite{agarcia-vela:06}, we assume that the pump laser acts
during an extremely short time and the system undergoes a perfect vertical
transition.  The Schr\"{o}dinger equation for the tetra-atomic system in both
the ground and excited states is solved using the variational method proposed in
Ref.~\onlinecite{pvillarreal:94}.  Then, the $X$-state wave function is used to
calculate the probability distributions (statistical weights) for all relevant
variables, while the eigenvalues from the $B$-state determine the VP energetics,
as in our previous work \cite{mlglez-mtnez:06a,mlglez-mtnez:08}.  Once the $X$-
and $B$-state quantum problems are solved, QCT initial conditions are calculated
following the algorithm proposed in Ref.~\onlinecite{gdelgadobarrio:97} (with
only a few minor modifications):
\begin{enumerate}
 \item a set of `quantum' numbers $\{l_1, l_2, j\}$ is selected using their
  respective distributions---checking it satisfies the triangle condition
  $\triangle(l_1, l_2, j)$ to ensure compatibility with $\boldsymbol{J} =
  \boldsymbol{0}$;

 \item the angles $\theta_1$, $\theta_2$, $\gamma$ are generated following the
  appropriate distributions, and $\phi$ calculated from
  Eq.~\eqref{eq:qct_cosgamma};

 \item the system of non-linear equations \eqref{eq:qct_li} and \eqref{eq:qct_j}
  is solved for the angular momenta $P_{\theta_1}$, $P_{\theta_2}$ and
  $P_{\phi}$;

 \item $r$ is generated from the calculated distribution, and the value of $j$
  used to calculate $P_r$ from Eq.~\eqref{eq:qct_HBC}.  Here,
  $\mathcal{H}_\mathrm{BC}$ is conveniently replaced with
  $\mathcal{E}^{B,v'}_\mathrm{BC}$, \emph{i.e.}~the eigenvalue corresponding to
  the ($B,v',j'=0$) state of BC \cite{pvillarreal:94}, and the sign of $P_r$
  randomly chosen;

 \item $R_1$ and $R_2$ are generated with the corresponding distribution, and
  approximate conjugate momenta calculated using
  \begin{equation}
   P^\mathrm{app}_{R_i} = \pm \sqrt{2 \mu_\mathrm{Rg_i,BC}
    \left( \mathcal{E}^{B,v'}_\mathrm{Rg_i,BC}
          - \frac{l^2_i}{2 \mu_\mathrm{Rg_i,BC} R^2_i}
          - V_\mathrm{Rg_i,BC} \right)},
   \label{eq:Papp}
  \end{equation}
  where $\mathcal{E}^{B,v'}_\mathrm{Rg_i,BC}$ is the eigenvalue corresponding to
  the Rg$_i$BC triatomic aggregate with BC in its ($B,v'$) state
  \cite{pvillarreal:94}, and the sign of the momentum is once again randomly
  chosen;

 \item finally, the energy associated to the non-diagonal kinetic term
  $\boldsymbol{P}_1 \cdotp \boldsymbol{P}_2$ and $V_\mathrm{Rg_1 Rg_2}$ is
  redistributed within the vdW modes by writing $P_{R_{1(2)}} =
  P^\mathrm{app}_{R_{1(2)}}  + \Delta_{1(2)}$, where $\Delta_1$ and $\Delta_2$
  are quantities to be determined.  Their calculation requires evaluating
  $\mathcal{H}^{\boldsymbol{J}=\boldsymbol{0}} =
  \mathcal{E}^{B,v'}$,  the energy of the tetra-atomic complex
  with Br$_2$ in the ($B,v'$) state, using the set of  values $\{ r, R_1,
  R_2, \theta_1,  \theta_2, \phi, P_r,  P^\mathrm{app}_{R_1},
  P^\mathrm{app}_{R_2}, P_{\theta_1},  P_{\theta_2},  P_\phi \}$, as well as the
  additional constrain $P^\mathrm{app}_{R_1}/P^\mathrm{app}_{R_2} =
  \Delta_1/\Delta_2$.
\end{enumerate}
Steps 1--6 are repeated until $N_\mathrm{tot}$ initial conditions are obtained.

\subsubsection{From classical magnitudes to observables}
\label{sec3:statistics}
We have recently discussed the statistical handling of QCT results for
comparison with experimental observables in the specific case of the VP of
triatomic vdW systems \cite{mlglez-mtnez:08}.  There is however one fundamental
difference between the process in tri- and tetra-atomic molecules, namely, the
possible formation in the latter of an intermediate complex.  For consistency,
the general methodology is only summarized here with emphasis on the changes
made in order to analyze the QCT results in our specific case.

When the molecule completely dissociates, $V_\mathrm{vdW}\rightarrow 0$ and both
$\boldsymbol{j}$ and $\mathcal{H}_\mathrm{BC}$ become integrals of motion with
the latter corresponding to the classical energy of the BC fragment,
$E_\mathrm{BC}$.  The final semi-classical ro-vibrational quantum numbers for
the $i$th dissociated trajectory, \emph{i.e.}~$v^\mathrm{f}_{c,i}$ and
$j^\mathrm{f}_{c,i}$, are then rigorously constant and are given by
\begin{eqnarray}
 v^\mathrm{f}_{c,i} &=& \frac{1}{2\pi\hbar}\oint{P_r dr}-\frac{1}{2}\nonumber\\
                    &=& \frac{\sqrt{2\mu_\mathrm{BC}}}{\pi\hbar}
                        \int_{r_\mathrm{min}}^{r_\mathrm{max}}
                      {\sqrt{E_{\mathrm{BC},i}-V_\mathrm{eff}}\;dr}-\frac{1}{2},
\label{eq:v}
\end{eqnarray}
where $V_\mathrm{eff}$ is the effective interaction potential of BC (including
the centrifugal term) and the closed integral is evaluated over one BC
vibrational period \cite{whmiller:74} ($r_\mathrm{min}$ and $r_\mathrm{max}$ are
the classical turning points), while
\begin{eqnarray}
 j^\mathrm{f}_{c,i} = \frac{1}{2}
     \left[\sqrt{1 + 4\left(\frac{\boldsymbol{j}_i}{\hbar}\right)^2} - 1\right],
\end{eqnarray}
which results from
$\boldsymbol{j}^2_i = j^\mathrm{f}_{c,i} (j^\mathrm{f}_{c,i} + 1) \hbar^2$
\cite{rnporter:76}.  Due to the weakness of the vdW interaction, the
BC vibrational state in the intermediate complex, $v^\mathrm{i}_{c,i}$, can be
estimated using Eq.~\eqref{eq:v}.  This requires replacing the final state
magnitudes with approximate values for the intermediate state, despite these
not being rigorously defined.

In what follows, we neglect the quantized nature of the rotational DOF as it has
a very high density of states and refer simply to channel
$\Delta v' = v^\mathrm{f} - v'$ by the associated final vibrational quantum
number, $v^\mathrm{f}$.

In the procedure known as \emph{histogram} or \emph{standard} binning (SB), the
probability density $P(M,v^\mathrm{f})$ that a given observable has the value
$M$ and the final BC vibrational state is $v^\mathrm{f}$ can be formally written
as
\begin{equation}
 P_\mathrm{SB}(M, v^\mathrm{f}) = \int d\mathbf{\Gamma}
         \rho(\mathbf{\Gamma}) \delta\left[M(\mathbf{\Gamma}) - M\right]
             \Xi\left[v^\mathrm{f}(\mathbf{\Gamma}); v^\mathrm{f}, 1\right],
\label{eq:Psb1}
\end{equation}
where $\rho(\mathbf{\Gamma})$ is the probability distribution of the initial
phase space state, $\mathbf{\Gamma}$, and
\begin{equation}
 \Xi(x; x^*, \Delta) = \frac{1}{\Delta}\Theta\left(x^* + \Delta/2 - x\right)
                       \Theta\left(x - x^* + \Delta/2\right).
\label{eq:xi}
\end{equation}
$\delta(x)$ and $\Theta(x)$ are respectively the Dirac and Heaviside functions.
$M(\mathbf{\Gamma})$ and $v^\mathrm{f}(\mathbf{\Gamma})$ are the final values of
the observable and the vibrational action in terms of $\mathbf{\Gamma}$.  It is
relatively easy to see from Eq.~\eqref{eq:xi} that $\Xi$ defines a square
barrier function of $x$, which equals $1/\Delta$ on $[x^* - \Delta/2, x^* +
\Delta/2]$ and 0 everywhere else. Also, $P_\mathrm{SB}$ in
Eq.~\eqref{eq:Psb1} is normalized to 1.

In practice, $P_\mathrm{SB}$ is estimated by means of the Monte-Carlo expression
\begin{eqnarray}
 P_\mathrm{SB}(M_k, v^\mathrm{f}) &\approx& \frac{1}{N_\mathrm{diss}}
  \sum^{N_\mathrm{diss}}_{i=1}
                 \Xi\left[M(\mathbf{\Gamma}_i); M_k, \alpha_k\right] \nonumber\\
  &&\times \Xi\left[v^\mathrm{f}(\mathbf{\Gamma}_i); v^\mathrm{f}, 1\right],
\label{eq:Psb2}
\end{eqnarray}
where $M_k$ is the $k$th midpoint in the $\{M\}^{K}_{k=1}$ set partition of the
interval $[M_\mathrm{min},M_\mathrm{max}]$, $\alpha_k = M_{k+1} - M_k$ and
$N_\mathrm{diss}$ is the total number of dissociated trajectories, which is
assumed to be large.

In the GW procedure one simply replaces the square barrier function
$\Xi\left[v^\mathrm{f}(\mathbf{\Gamma}); v^\mathrm{f}, 1\right]$ in
Eq.~\eqref{eq:Psb2} with the Gaussian
\begin{equation}
 g\left[v^\mathrm{f}(\mathbf{\Gamma}); v^\mathrm{f}, \epsilon\right] =
         \frac{\exp{\left\{-\left[v^\mathrm{f}\left(\mathbf{\Gamma}\right) -
                  v^\mathrm{f}\right]^2/\epsilon^2\right\}}}{\pi^{1/2}\epsilon},
\label{eq:gw}
\end{equation}
in which $\epsilon$ is usually kept at 0.05.  Hence, $P_\mathrm{GW}$ is simply
\begin{eqnarray}
 P_\mathrm{GW}(M_k, v^\mathrm{f}) &\approx& \frac{1}{N_\mathrm{diss}}
  \sum_{i=1}^{N_\mathrm{diss}}
                 \Xi\left[M(\mathbf{\Gamma}_i); M_k, \alpha_k\right] \nonumber\\
  &&\times g\left[v^\mathrm{f}(\mathbf{\Gamma}_i); v^\mathrm{f},\epsilon\right].
\label{eq:Pgw}
\end{eqnarray}
In the tetra-atomic case, variations of these general expressions can describe
either the intermediate or final states. To do this, it is sufficient to use the
appropriate weight function for the desired state for the second $\Xi$ function
in Eq.~\eqref{eq:Psb2} and the Gaussian function in Eq.~\eqref{eq:gw}
respectively. We must note that the weight for the intermediate state is
calculated as the product of the  weights corresponding to the semi-classical
vibrational level of BC and that of the intermediate vdW complex: $w^\mathrm{i}
= w^\mathrm{i}_1\left(v^\mathrm{i}_c\right)
w^\mathrm{i}_2\left(n^\mathrm{i}_c\right)$, where $w$ refers to either $\Xi$
or $g$.  This could dramatically increase the number of trajectories necessary
for convergence.  Moreover, although Eq.~\eqref{eq:v} provides a simple means to
estimate the semi-classical analogue for the vibrational state of BC, that of
the vdW complex is much more difficult to evaluate.  A simpler approach, known
as the 1GB procedure, uses the total ro-vibrational energy rather than that of
the individual states to calculate the corresponding weight: $w^\mathrm{i} =
w^\mathrm{i}\left(E^{B,v^\mathrm{i}_c}_\mathrm{Rg_iBC}\right)$.  Both the
$\mathcal{E}^{B,v'}_\mathrm{BC}$ and $\mathcal{E}^{B,v'}_\mathrm{Rg_i,BC}$
eigenvalues are involved in this calculation, see Sec.~\ref{sec3:qct_ic.theory}.
The method was proposed by Czak\'o and Bowman \cite{gczako:09} on the basis of
rather intuitive arguments and was later theoretically validated by Bonnet and
Espinosa-Garc\'{\i}a \cite{lbonnet:10}.  In the particular case when the first
and second dissociation steps are statistically independent, the total weight
can be calculated as the product $w = w^\mathrm{i} w^\mathrm{f}$.  In addition,
mixed strategies may be tested by simply using $w = w^\mathrm{i(f)}$, hence
calculating the desired state distributions independently of the
final/intermediate vibrational state.

The time evolution of the populations of the various complexes can be obtained
by small modifications to the formulae determining survival probabilities,
\emph{i.e.}\ the probability that a given complex has not dissociated at time
$t$, which read \cite{mlglez-mtnez:08}
\begin{eqnarray}
 P_{\mathrm{SB}}(t) &=& 1-\frac{1}{N_\mathrm{diss}}\sum_{i=1}^{N_\mathrm{diss}}
  \Theta(t-t_i),\nonumber\\
 P_{\mathrm{GW}}(t) &=& 1 - \frac{1}{\sum_{i=1}^{N_\mathrm{diss}}g_i}
  \sum_{i=1}^{N_\mathrm{diss}}g_i\Theta(t-t_i),
\label{eq:PSt}
\end{eqnarray}
with $g_i = \sum_{\forall v} g(v(\mathbf{\Gamma}_i); v, \epsilon)$ and $t_i$
the dissociation time for the $i$th trajectory.  In the particular case of a
Rg$_2$BC cluster, for which $t_1$ and $t_2$ are the dissociation times for the
first and second Rg atoms, the different populations are explicitly obtained
as: (1) $P_4(t) = P(t_1)$,  the decay of the parent molecule; (2) $P_3(t) =
P(t_2) - P(t_1)$, time evolution of the intermediate RgBC population; and (3)
$P_2(t) = 1 - P(t_2)$, formation of the BC fragment.  Moreover, (4) $P'_3(t) =
P(t_2 - t_1)$ directly represents the decay of the RgBC complex.  The reasons
for the subscripts will become apparent in Section~\ref{sec2:kinetics}.

\subsubsection{Simulation details}
\label{sec3:qct_simulation}
Batches of $N_\mathrm{tot} = 5 \times 10^5$ trajectories were propagated using
an adaptive-stepsize Bulirsch-Stoer method \cite{whpress:book07}.  Each
trajectory was followed until one of two conditions was fulfilled: (1) the two
Ne atoms dissociated, \emph{i.e.}~$R_{1(2)} \ge R^\mathrm{QCT}_\mathrm{diss} =
14$~\AA; or (2) the propagation time $t = T_\mathrm{max} = 1200$~ps.
Dissociation is said to occur when $R_\emph{1(2)} = 14$, the distance at which
the vdW interactions become negligible.  The integration parameters were
adjusted so that the maximum error in total energy did not exceed $\Delta E =
10^{-5}$~cm$^{-1}$, \emph{i.e.}~less than 10$^{-8}\,E$.  With this choice of
parameters all trajectories dissociated at least one Ne atom while over
88--99\%, depending on $v'$, completely fragmented.  A typical trajectory
requires less than 1 second of CPU time on an
Intel\textsuperscript\textregistered Core\textsuperscript\texttrademark I7 Q720
(6M~Cache, 1.60~GHz) processor.

\subsection{Cartesian coupled coherent states}
\label{sec2:cccs}
The CCCS method is a trajectory-based quantum dynamics technique designed to be
similar to classical trajectory simulations.  The main focus of this paper is
the QCT predictions and thus CCCS results are mainly presented to corroborate
the conclusions and the quality of the QCT results.  Consequently, we will only
give a brief outline of CCCS in order to aid comprehension of the results that
are included.  The interested reader is referred to Ref.~\onlinecite{skreed:11a}
which not only describes in detail our previous work on the VP of NeBr$_2$ but
also outlines the extension to larger clusters.  A more detailed CCCS study of
Ne$_2$Br$_2$ will be published in due course \cite{skreed:12x}.

The QCT method described hitherto is concerned with the point-like nuclei that
make up the Ne$_2$Br$_2$ cluster whereas the CCCS method allows us to study the
time evolution of the associated quantum mechanical wave function.  The CCCS
method expands the wave function using a basis of coherent states (CS).  In
CCCS, the CS are Gaussian-shaped wave packets that describe both the position
and momentum for each Cartesian DOF of each atom.  Thus for Ne$_2$Br$_2$, each
CS has 24 dimensions.  The CS move on the same PES, and according to the same
equations as the nuclei in QCT.  However, the potential energy for the CS is the
convolution of the CS and the PES.  In contrast, within the QCT method, the
potential energy of a given configuration is simply the value of the PES at that
point in configuration space.  In CCCS, the convolution may be done in advance
of the simulations and it gives the so-called \emph{averaged} or
\emph{re-ordered} PES (or equivalently, Hamiltonian) upon which the centers of
the CS move.

The amplitudes evolve with time according to an expression derived from
Schr\"odinger's equation and which depends upon the fact that the basis
functions overlap.  This overlap couples the amplitudes of the basis functions
together.  If the basis functions are separated by a sufficiently large distance
in the 24-dimensional phase space, they cease to overlap and become decoupled.
Once all the basis functions have decoupled, the CCCS method is essentially a
semi-classical method.  However, even in its semi-classical limit, the CCCS
technique takes into account the majority of zero point energy effects.

\subsubsection{Simulation details}
\label{sec3:cccs_simulation}
The initial coordinates for the basis functions are determined from a set of
bond lengths and momenta so that the total linear and angular momenta of the
cluster are zero.  For Br--Br, these were chosen from the phase-space trajectory
at the energy expectation value of the isolated molecule.  In the case of Br--Ne
and Ne--Ne, they were chosen randomly from the ground state wave functions of
the bonds.  Initially, each basis consisted of 400 basis functions and the
results have been averaged over 10 different basis sets per Br$_2$ vibrational
level.

As CCCS simulations are considerably more expensive than QCT calculations, a Ne
atom is said to have dissociated once the mean distance between it and the
Br$_2$ molecule exceeds $R^\mathrm{CCCS}_\mathrm{diss} = 10$~\AA.  Similarly,
the second stage dissociation is said to occur when the distance between Br$_2$
and the remaining Ne exceeds the same cutoff.

In our previous work~\cite{skreed:11a}, we found that the basis functions in the
simulations of the VP of NeBr$_2$ remained coupled until the clusters completely
dissociate.  In the current work, however, the basis functions decouple much
more quickly due to the increased dimension of phase space.  Obviously, the time
at which each basis function decouples varies greatly but it is typically less
than 10~ps: the mean time in one simulation was about 6~ps.  As will be
discussed later, this leaves comparatively few independent basis functions to
describe the dissociating wave function and leads to increased lifetimes. 

The calculated lifetimes can be improved by spawning a secondary simulation each
time a basis function permanently decouples in which the decoupled basis
function is expanded on a new basis of 50 CS.  The calculation of the lifetimes
is thus a two-part process: first we perform a simulation of the whole wave
function expanded on the basis of 400 CS; and then we perform 400 secondary
simulations each with 50 basis functions describing one of the basis functions
of the original simulation starting from when that basis function decoupled.  We
found that decoupling invariably occurs before dissociation which makes the
combination of the dissociation curves from 400 secondary simulations easier.

The time required to perform the CCCS calculations depends upon the length of
time for which the basis functions remain coupled and increases nonlinearly with
the number of basis functions.  As an example, the mean time for $v = 17$ for
100~ps ($10^5$ steps), using 400 basis functions was 3.2 days.  The mean time
for the re-expansions however was 1.3 hours per simulation on
Intel\textsuperscript\textregistered Xeon\textsuperscript\textregistered~
`Woodcrest' processors.

\section{Results and Discussion}
\label{sec:results}
\subsection{Kinetics, lifetimes}
\label{sec2:kinetics}
\subsubsection{QCT}
\label{sec3:kinetics_qct}
PSTT establishes that certain structures in classical phase space (resonant
islands, tori, cantori\ldots)~act as inter- and intramolecular bottlenecks to
the diffusion of trajectories from regions defining `reactants' and `products'
\cite{mjdavis:86,skgray:87,regillilan:91}.  A rigorous (classical) kinetic
mechanism must therefore reflect the details of these partial obstacles.  In
addition, a complete rationalization of the energy-dependence of decay rates
requires the stability of periodic orbits and the evolution of bifurcations in
the phase-space portrait to be addressed in depth.  While this is possible and
has already been carried out for simple 2-DOF models of molecular systems,
including vdW complexes \cite{mjdavis:85,skgray:86,mzhao:92c,rprosmiti:02c,
aagranovsky:98}, its extension to more DOF introduces serious technical
difficulties \cite{regillilan:91} and constitutes in itself an active field of
research.  In fact, further below we use some of these results and a 2-DOF model
of NeBr$_2$ to better understand the relation between product states and
dissociation lifetimes, as well as certain features in the time-evolution of the
intermediate complex.

Analyzing the phase-space structure of Ne$_2$Br$_2$, or even NeBr$_2$, is
however far from our objective here and, at this point, we take advantage of one
important experimental conclusion: At least for low $v'$ levels, a simple
sequential kinetic mechanism (Section~\ref{sec2:VP_Rg1Rg2BC}) provides an
excellent fit to the delay scans (Figs.~7 and 8 in Ref.~\onlinecite{jmpio:10}).
This is consistent with several results from our QCT calculations: (1) once the
Ne$_2$Br$_2$ complex decayed into NeBr$_2$, the only process observed was the
fragmentation of the resulting triatomic into Ne and Br$_2$ products; (2) less
than about 0.05\% of all trajectories dissociated via the concerted mechanism
(subset of all trajectories for which both Ne atoms dissociated within one
Br$_2$ vibrational period, excluding those predicted by the sequential mechanism
to dissociate within that time interval \footnote{In order to further
distinguish between $C_1$ and $C_2$ events in Scheme~\ref{sch:VP_Rg1Rg2BC}, the
Ne--Ne energy was checked for compatibility with formation of a Ne$_2$
molecule.}); and (3) no significant statistical correlation was found between
the variables defining the intermediate complex and $t_2 - t_1$.  The latter
confirms, at least from the kinetics viewpoint, that the second dissociation
step is nearly statistically independent on the first \footnote{One may consider
the case where the initial dissociation step leads---for a significant part of
the ensemble---to states `trapped' behind large centrifugal barriers,
\emph{i.e.}~orbiting resonances.  These would act as effective bottlenecks to
the dissociation of the triatomic intermediate complex and result in a slightly
more complicated kinetic mechanism, as in Ref.~\onlinecite{agarcia-vela:96}.  In
such cases, the rotational angular momentum of the intermediate complex would
result statistically correlated with $t^i_2 - t^i_1$.}.  All of these
observations agree with the fact that: (4) EQM calculations on the VP of
NeBr$_2$ show that direct predissociation dominates within the vibrational range
explored here \cite{agarcia-vela:06}.

\begin{figure*}[!t]
 \begin{center}
  \includegraphics[width=164mm]{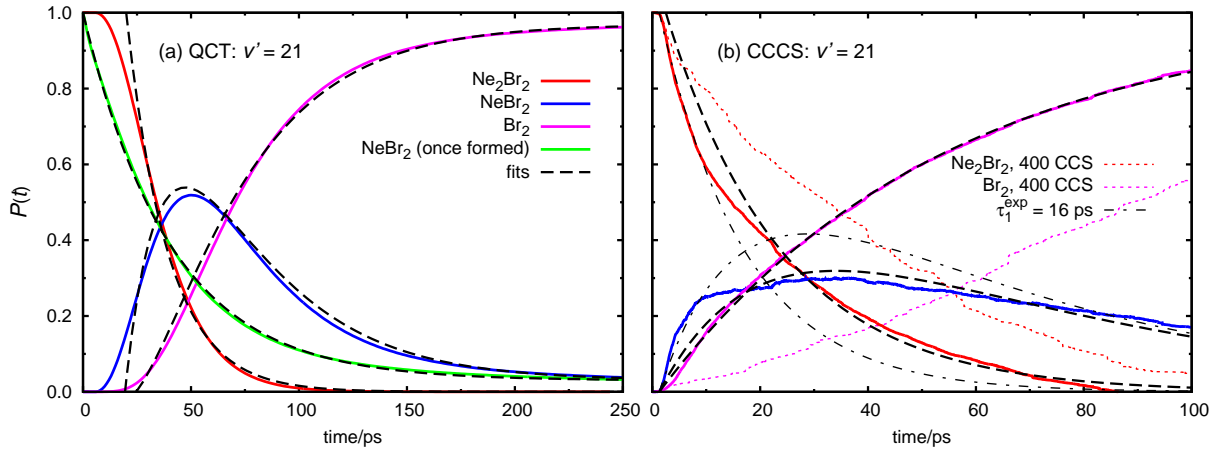}
  \caption{(Color online) Time evolution of the population of all complexes
   involved in the vibrational predissociation of Ne$_2$Br$_2$($B$, $v' = 21$):
   (a) QCT; and (b) CCCS calculations.
   \label{fig:survprob}}
 \end{center}
\end{figure*}

The proposed kinetic mechanism is schematically
\begin{equation}
 N_4\stackrel{\mathrm{\tau_1}}{\longrightarrow}N_3
                                 \stackrel{\mathrm{\tau_2}}{\longrightarrow}N_2,
\label{eq:kmec1}
\end{equation}
where $N_4$, $N_3$ and $N_2$ stand for the Ne$_2$Br$_2$, NeBr$_2$ and Br$_2$
populations respectively, while $\tau_1$ and $\tau_2$ are the corresponding
lifetimes.  Mathematically, Eq.~\eqref{eq:kmec1} can be written as
\begin{eqnarray}
 \mathrm{d}N_4/\mathrm{d}t &=& -k_1 N_4, \nonumber\\
 \mathrm{d}N_3/\mathrm{d}t &=& k_1 N_4 - k_2 N_3, \nonumber\\
 \mathrm{d}N_2/\mathrm{d}t &=& k_2 N_3,
 \label{eq:kmec2}
\end{eqnarray}
with $k_i = \tau^{-1}_i$, and its solution
\begin{eqnarray}
 N_4(t) &=& N_4(0) \exp{(-k_1 t)}, \nonumber\\
 N_3(t) &=& \left[N_3(0) - \frac{k_1 N_4(0)}{k_2 - k_1}\right] \exp{(-k_2 t)}
          +\frac{k_1 N_4(t)}{k_2 - k_1},\nonumber\\
 N_2(t) &=& N_\mathrm{tot} - N_4(t) - N_3(t),
\label{eq:kmec3}
\end{eqnarray}
where $N_4(0) = N_\mathrm{tot}$ and $N_3(0) = N_2(0) = 0$.  Given the dependence
on $\tau_1$ and $\tau_2$ of Eqs.~\eqref{eq:kmec3}, various schemes to extract
the lifetimes from survival probability curves become possible.  One can: (1, 2)
first obtain $\tau_1$ from fits to Ne$_2$Br$_2$ decay curves and use this value
to get $\tau_2$ from the time dependence of NeBr$_2$, or alternatively, Br$_2$
populations; or (3, 4) directly infer both values from the evolution of NeBr$_2$
or Br$_2$ populations.  Alternatively, taking into account that the individual
dissociation times $t_1$ and $t_2$ are nearly uncorrelated, one may: (5)
directly extract the value of $\tau_2$ by fitting the probability of complete
dissociation after a time $t_{21} = t_2 - t_1$ (fragmentation of the triatomic
once it is formed) to a single exponential decay.  This is implicit in our
kinetic mechanism and readily seen by adding the first two equations in
\eqref{eq:kmec2}.  Schemes (1)--(4) are somewhat similar to the experimental
freedom of determining the lifetimes by measuring the time dependence of the
Ne$_2$Br$_2$/NeBr$_2$ signal in a given Br$_2$ vibrational state, referred to as
`disappearance' of the parent complex, and that of the NeBr$_2$/Br$_2$ signal in
the vibrational states resulting from dissociation, the `appearance' of the
daughter molecule, etc.  Depending on the signal monitored, different values are
obtained in the experiments (and EQM), which provide a measure of IVR
\cite{jacabrera:05,agarcia-vela:06,jmpio:10}.  However, QCT will yield the same
lifetimes no matter what scheme is used (though numerical fitting and
convergence will in practice result in small differences).  This is easily
understood for according to our criteria, the disappearance of Ne$_2$Br$_2$
exactly matches the appearance of NeBr$_2$, etc.  As an example, the relevant
probability curves, \emph{i.e.}~$P_i(t) = N_i(t)/N_\mathrm{tot}$, $i =$~4--2
and $P'_3(t)$, corresponding to $v' = 21$ are shown in panel (a) of
Fig.~\ref{fig:survprob}, together with the respective fits.
We have verified that the various schemes yield similar results, and
experimental and theoretical lifetimes are summarized in
Table~\ref{tab:lifetimes}.
\begin{table}[!b]
 \begin{center}
  \caption{Experimental \cite{jmpio:10} and theoretical---this work and MDQT
   \cite{bmiguel:01}---lifetimes (in ps) in the vibrational predissociation of
   Ne$_2$Br$_2$($B, v' =$~16--23).
   \label{tab:lifetimes}}
  \begin{tabular}{cccccccccccc}
   \hline\hline
        &                  & \multicolumn{2}{c}{Experiment} && \multicolumn{2}{c}{QCT} && \multicolumn{2}{c}{CCCS} && MDQT     \\ \cline{3-4}\cline{6-7}\cline{9-10}\cline{12-12}
   $v'$ & Species($v$)     & $\tau_1$ & $\tau_2$            && $\tau_1$ & $\tau_2$     && $\tau_1$ & $\tau_2$      && $\tau_1$ \\ \hline
   16   & Ne$_2$Br$_2$(16) &          &                     && 62.1     & 134.0        &&          &               &&          \\ \hline
   17   & Ne$_2$Br$_2$(17) & 32$\pm$3 &                     && 48.5     & 115.9        && 44.0     & 98.9          && 29       \\
        & NeBr$_2$(16)     & 30$\pm$3 & 88$\pm$3            &&          &              &&          &               &&          \\
        & Br$_2$(15)       & 31$\pm$2 & 82$\pm$3            &&          &              &&          &               &&          \\ \hline
   18   & Ne$_2$Br$_2$(18) & 28$\pm$3 &                     && 33.8     & 89.6         &&          &               &&          \\
        & NeBr$_2$(17)     & 27$\pm$5 & 58$\pm$5            &&          &              &&          &               &&          \\
        & Br$_2$(16)       & 28$\pm$3 & 55$\pm$4            &&          &              &&          &               &&          \\ \hline
   19   & Ne$_2$Br$_2$(19) &          &                     && 28.0     & 66.5         &&          &               && 24       \\ \hline
   20   & Ne$_2$Br$_2$(20) &          &                     && 23.9     & 51.1         &&          &               && 19       \\ \hline
   21   & Ne$_2$Br$_2$(21) & 16$\pm$3 &                     && 19.3     & 40.5         && 21.7     & 53.1          && 16       \\
        & NeBr$_2$(20)     & 14$\pm$5 & 30$\pm$2            &&          &              &&          &               &&          \\
        & NeBr$_2$(19)     & 17$\pm$4 & 54$\pm$3            &&          &              &&          &               &&          \\
        & Br$_2$(19)       & 16$\pm$2 & 29$\pm$2            &&          &              &&          &               &&          \\
        & Br$_2$(18)       & 16$\pm$2 & 47$\pm$2            &&          &              &&          &               &&          \\ \hline
   22   & Ne$_2$Br$_2$(22) &          &                     && 16.2     & 34.5         &&          &               && 13       \\ \hline
   23   & Ne$_2$Br$_2$(23) &          &                     && 14.1     & 28.9         && 15.3     & 39.4          &&          \\
   \hline\hline
  \end{tabular}
 \end{center}
\end{table}
In the latter, column `$v'$' corresponds to the vibrational level in the $B$
electronic state to which the Ne$_2$Br$_2$ molecule is excited by the laser
pulse.  Column `Species($v$)' then corresponds to the molecular product (in the
specific vibrational state $v$) that is monitored in the experiment.  `$\tau_1$'
and `$\tau_2$' are the lifetimes of  Ne$_2$Br$_2$ and  NeBr$_2$.  In particular,
as seen in Eqs.~\eqref{eq:kmec3}, the fragmentation of Ne$_2$Br$_2$ depends only
on $k_1$, which is why no $\tau_2$ is reported in the experiment for this
molecule.  As discussed above, the theoretical methods yield nearly identical
results independently on the probability curve used for the fitting, and a
single pair of values $\tau_1$, $\tau_2$ is thus reported for the system (note
that no $\tau_2$ value was reported in Ref.~\onlinecite{bmiguel:01}).

Simple inspection of panel (a) in Fig.~\ref{fig:survprob} shows that the
agreement between our fits and QCT calculations is very good, which confirms the
accuracy of the kinetic mechanism proposed.  This is the general trend for all
vibrational levels explored here.  Even so, in most cases, small discrepancies
occur at short and large times.  From the comparison of the classical survival
probability curves corresponding to the decay of the Ne$_2$Br$_2$, red curve,
and the fragmentation of the intermediate NeBr$_2$, green curve, the nature of
the plateau usually observed at short times becomes rather clear.  It is simply
an artifact resulting from both the classical description of the process and the
way initial conditions are sampled.  More explicitly: there is a minimum time
for the gradual classical energy transfer from the Br$_2$ vibrational mode to
dissociate one Ne atom which, in the case of the first decay, has been located
close to the vdW minima at the belt-like configuration.  In contrast, a quantum
vibrational transition may immediately release enough energy for dissociation to
occur, as readily seen in CCCS curves from panel (b).  The fact that no such
feature is observed in the second QCT step further confirms this reasoning for
once the first atom is lost, the dynamics would sample more evenly the available
phase space and have placed the second Ne atom arbitrarily close to
$R_\mathrm{diss}$.  One may say that this apparent non-exponential behavior (in
the fragmentation kinetics) mainly arises from initial state selection
\cite{dlbunker:73}.

The reasons for the discrepancies at larger times are more complex and better
understood using a 2-DOF model for the VP of NeBr$_2$ (with $\theta=\pi/2$).
Analysis of the SOS shows a 1:10 nonlinear resonance (for $v' = 21$) surrounding
the stable central region around the point $(R = R_\mathrm{eq}, P_R = 0)$.
This resonance occupy a significant proportion of the available phase space.
The extent to which such intramolecular bottlenecks affect the overall time
evolution of the classical ensemble depends on the proportion of initial
conditions lying inside or relatively close to their boundaries.  Actually,
early work on PSTT showed that kinetic mechanisms can be conveniently modified
to account for such behavior \cite{mjdavis:86}.  In our case, only a relatively
low percent of trajectories, about 0.3\% for $v' = 21$ in
Fig.~\ref{fig:survprob}, show a strong non-exponential behavior, \emph{always}
after the dissociation of the first Ne atom.  This is most likely due to
remaining lower-dimensionality tori, which have been recently demonstrated to
play a significant role as bottlenecks between diffusive and statistical
behavior in systems with more than two DOF \cite{rpaskaukas:08}.

Given all of the above, the actual equations used to produce the fits shown in
Fig.~\ref{fig:survprob} are in fact slightly modified versions of
Eqs.~\eqref{eq:kmec3} which account for both sources of non-exponential
behavior.  That at low times is `avoided' by fitting from $t^{v'}_0 > 0$, while
the long-time behavior is modeled by adding a given constant (which equals the
asymptotic proportion of non-dissociated trajectories at each $v'$).  We should
note that, after adding this constant, all coefficients need to be adequately
modified to recover the correct behavior at $t = 0$.

\subsubsection{CCCS}
\label{sec3:kinetics_cccs}
The survival probabilities for the different complexes for $v' = 21$, as 
calculated using CCCS, are shown in panel (b) of Fig.~\ref{fig:survprob}.  In
contrast to panel (a): (1) dotted lines (labeled `400 CCS') correspond to the
initial CCS simulations with 400 basis functions, the re-expansion of which
gives the main curves; (2) the survival probability of NeBr$_2$ once it is
formed from the Ne$_2$Br$_2$ cluster is not presented, as this time is
ill-defined in the CCCS calculations; and (3) dot-dashed lines (labeled
`$\tau^\mathrm{exp}_1 = 16$~ps') uses the experimental value \cite{jmpio:10} to
model the time evolution at short times.  The fits are obtained using the same
sequential kinetic mechanism as for QCT, and the lifetimes given in
Table~\ref{tab:lifetimes} are from fitting the modified versions of
Eqs.~\eqref{eq:kmec3} to the CCCS data, also necessary here, mainly to account
for those basis functions that do not dissociate.

Theoretically, CCCS will provide a quantum mechanical description as long as the
basis functions remain coupled, which ideally should be for a time comparable to
the process of interest.  As in our previous study of the NeBr$_2$ system
\cite{skreed:11a}, CCCS removes the nonphysical plateau shown by QCT at short
times, thus accurately reproducing the initial quantum dynamics.  However, it
becomes harder to fit the results of CCCS to a simple model of sequential
dissociation at longer times.  This could be because quantum mechanics makes
non-sequential dissociation more likely.  Previously, non-sequential IVR-EC
contributions to dissociation were found to be important in MDQT simulations for
$v > 14$ where they account for one fifth of all dissociative trajectories
\cite{bmiguel:01}.  Also, experiments \cite{jmpio:10} seem to indicate that
complicated non-sequential mechanisms are important for $v > 19$.  However,
it would be premature to claim that the current CCCS simulations prove the
pertinence of non-sequential mechanisms.  Although our re-expansion technique
(Section~\ref{sec3:cccs_simulation}) greatly increases the time scale at which
CCCS works, it is still below the range of hundreds of picoseconds required for
the description of the whole fragmentation process.  At such long time scales,
CCCS works as a semi-classical technique running largely uncoupled CS on the
averaged potential.  It accounts for zero point energy effects but not for the
full quantum coupling in phase space.  We are currently working on strategies to
further extend the coupling time between the CS in the quantum simulations.
More detailed (and costly) investigations are necessary and the results will be
published elsewhere \cite{skreed:12x}.

In general, both QCT and CCCS predictions are in very good agreement with each
other, thus CCCS confirming the good quality of much simpler and less expensive
QCT results.  Both methods are also in good agreement with the experimental
lifetimes, over the whole range of initial vibrational states.  However,
although providing a near-quantitative description of the kinetics, they seem to
consistently yield larger than observed values.  In the case of QCT, this may be
intrinsic to the methodology and the existence of low-dimensional tori in
classical phase space.  Additional CCCS calculations are needed in order to
verify whether this is inherent to the method or due to, \emph{e.g.}\ an
inadequate representation of the PES or the wave function.

\subsection{PSTT, SB, GW and QCT dissociation times}
\label{sec2:PSTT+SB+GW+tdiss}
Based on PSTT, it seems reasonable to expect that trajectories with the highest
Gaussian weights, which contribute the most to the GW curves
and correspond to vibrational actions the closest to integer values, do not
necessarily explore the same regions in phase space than the whole ensemble and
could therefore undergo different kinetics.  Hence, at least in principle,
applying the SB or GW methods can produce different time evolution probability
curves (Section~\ref{sec3:statistics}) and correspondingly, result in different
lifetimes.  In turn, GW results should be the closest to the experimental values
for this method effectively eliminates trajectories which finish in nonphysical
vibrational channels.  Quite surprisingly, no significant differences between SB
and GW lifetimes were found in our previous calculations on the VP of RgBr$_2$
(Rg = He, Ne, Ar) molecules (see Fig.~1 in Ref.~\onlinecite{mlglez-mtnez:08}),
where several possible reasons were proposed.

In the VP of Ne$_2$Br$_2$ studied here, the predictive capabilities of GW can be
additionally tested for the experiment provides detailed information on the
kinetics associated to different vibrational channels for $v' = 21$ (see
Table~\ref{tab:lifetimes}).  Unfortunately, once again, our calculations seem to
corroborate that both the SB and GW methods result in nearly identical
lifetimes, with no significant variation between the various vibrational
channels.  In other words: the GW procedure was found to be unable to provide
vibrational resolution in the dissociation lifetimes.  In the rest of this
section, we will show that 2- and 3-DOF models for the VP of NeBr$_2$ are very
helpful in the detailed analysis of this issue.

\begin{figure}[!t]
  \includegraphics[width=81mm]{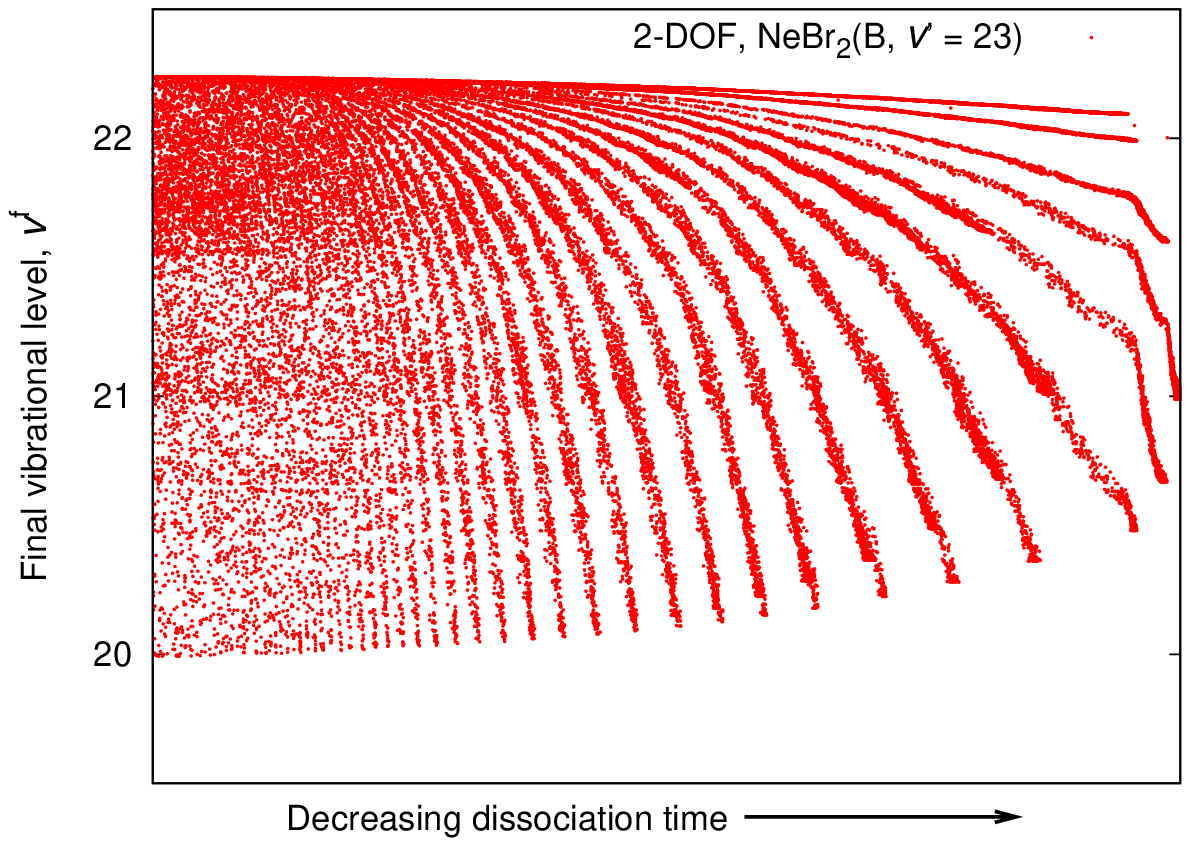}
  \includegraphics[width=81mm]{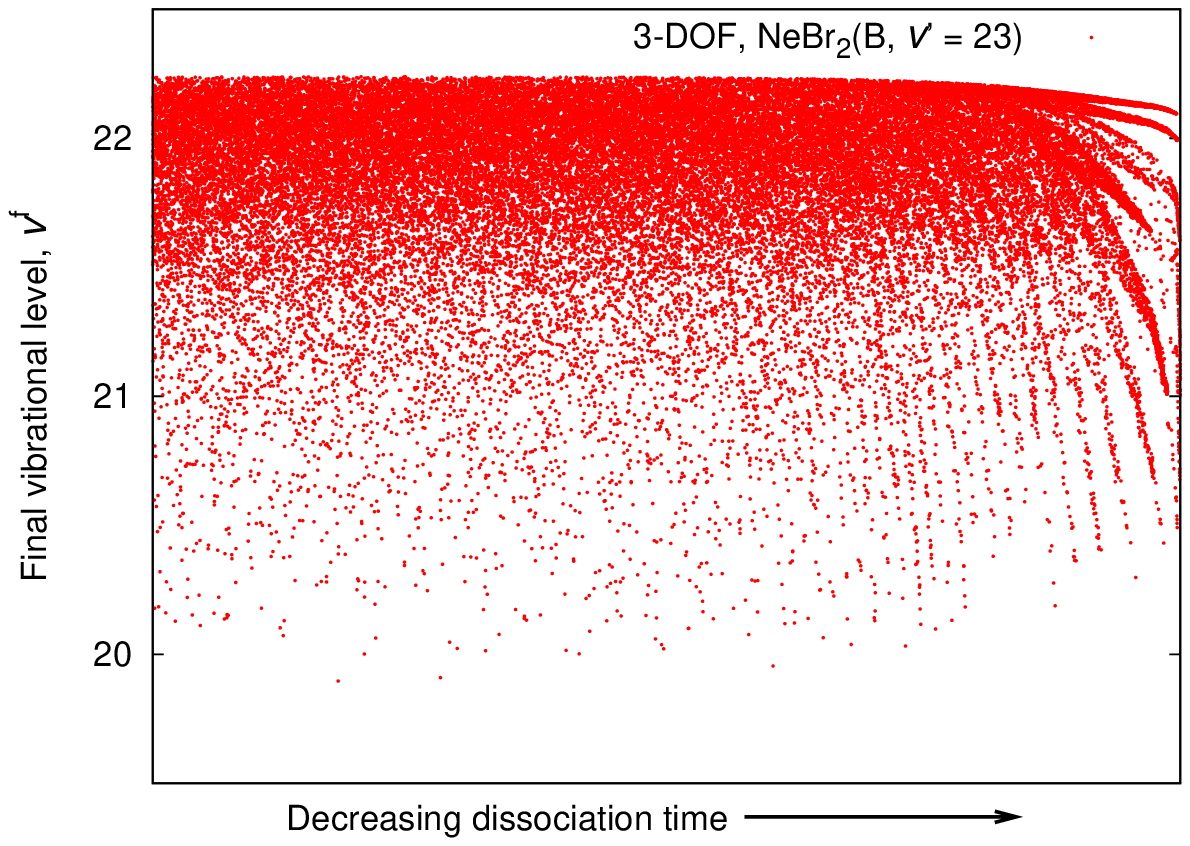}
  \caption{(Color online) Dependence between dissociation lifetimes and final
   vibrational states in the vibrational predissociation of NeBr$_2$ from $v' =
   23$: (a) 2-DOF; (b) 3-DOF model.
   \label{fig:tdissvf_23dof_v23}}
\end{figure}

The relation between trajectory lifetimes and final states is rather intricate,
as is apparent from Fig.~\ref{fig:tdissvf_23dof_v23}.  Panel (a) shows how final
vibrational states relate to the individual dissociation times for an ensemble
of $5 \times 10^4$ trajectories in the 2-DOF model.  As readily seen, there is a
relative insensitivity between these magnitudes for trajectories with longer to
medium lifetimes, while a specific pattern arises for shorter dissociation
times.  These finger-like structures can be understood by analyzing the VP
process in phase space, in the spirit of PSTT.  For simplicity we will only
introduce here the essential elements of the methodology that are required by
our interpretation.  Further information can be found in the seminal work of
Davis \emph{et al.}, most notably in Ref.~\onlinecite{mjdavis:86}.

\begin{figure}[!t]
 \begin{center}
  \includegraphics[width=85mm]{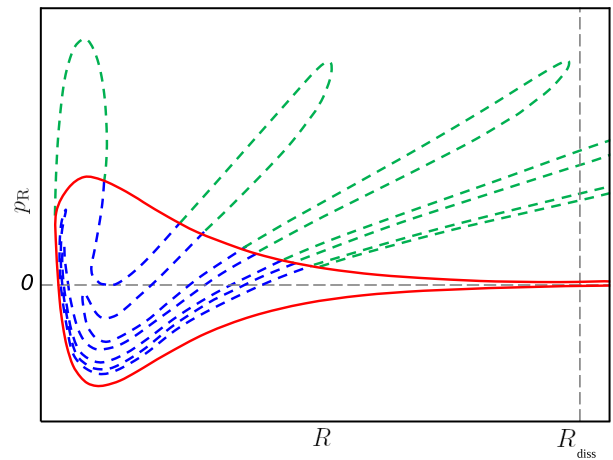}
  \caption{(Color online) Schematic view of the main structures in the SOS of a
   2-DOF model for the VP of a generic triatomic: separatrix (red, continuous
   line), outgoing (green, dashed line) and incoming (blue, dashed  line)
   fluxes.
   \label{diag:SOS}}
 \end{center}
\end{figure}

Figure~\ref{diag:SOS} depicts the main structures in the SOS of a 2-DOF model
for a generic triatomic.  The SOS is constructed by the procedure described in
Section~\ref{sec2:qct} and the molecule is assumed to dissociate if
$R \ge R_\mathrm{diss}$.  The \emph{separatrix} (continuous red line) encloses
the interaction region and constitutes the intramolecular bottleneck to
dissociation.  The fingers (green/blue dashed lines) limit the region available
to the outgoing/incoming flux.  All dissociating trajectories will intersect the
region enclosed by the outgoing fingers (dashed, green) and the separatrix,
once only.  Thus, after leaving the interaction region, any given trajectory
will inevitably dissociate and consecutive intersections will lie in different
fingers as $R(t)$ increases monotonically.  All that rests now is to notice
that trajectories closer to the fingers' `tips', \emph{i.e.}~those with the
largest $P_R$, correspond to the smallest $v$ available (and vice versa).  The
previous follows from the fact that the total energy $E$ is fixed in the
ensemble.  At last, the existence of finger-like structures like those in
Fig.~\ref{fig:tdissvf_23dof_v23} will strongly depend on the characteristics of
the dynamics within the interaction region.  If completely stochastic, initial
conditions will be effectively `forgotten' and trajectories will access the
outgoing fingers after exploring the interaction region during a random
propagation time.  No correlation between dissociation times and final
vibrational states will be observed.  On the other hand, if intermolecular
couplings are not strong enough, a subset of trajectories whose initial
conditions lie within a specific region inside the separatrix will directly
accesses the outgoing fingers. These trajectories will therefore dissociate
rapidly and a pattern will arise.  Following this line of reasoning, it is not
difficult to see why the structures in panel (a) of
Fig.~\ref{fig:tdissvf_23dof_v23} have precisely their shape.  Panel (b) in the
same figure shows results for the 3-DOF model of NeBr$_2$.  The similarities
between the 2- and 3-DOF models are evident, although the finger-like structures
get increasingly blurred the larger the number of DOF.  The number of such
finger-like structures depends on the kinetic energy available to dissociation,
hence decreasing with increasing vibrational excitation.  Additional
calculations show that all structures completely disappear for energies above
the linearization threshold (Br--Ne--Br configuration barrier).

\subsection{Product state distributions}
\label{sec2:pdistros}
Ro-vibrational product state distributions were calculated from the ensemble of
quasi-classical trajectories as described in Section~\ref{sec3:statistics}.  As
many relevant features of these observables strongly depend on the energetics of
the VP process, details of the latter are summarized in
Table~\ref{tab:energetics}: (1) the vibrational energy gap associated to the
$\Delta v' = -1$ channel, $\Delta\mathcal{E}^{B,v'-1}_\mathrm{Br_2} =
\mathcal{E}^{B,v'}_\mathrm{Br_2} - \mathcal{E}^{B,v'-1}_\mathrm{Br_2}$; (2, 3)
NeBr$_2$ \cite{mnejad-sattari:97} and Ne$_2$Br$_2$ binding energies,
$D^{\mathrm{NeBr_2}(B,v')}_0 = -\mathcal{E}^{B,v'}_\mathrm{NeBr_2}$ and
$D^{B,v'}_0 = -\mathcal{E}^{B,v'}$; and (4, 5) the energies available after
dissociation of the first and second Ne atoms via, respectively, the $\Delta v'
= -1$,~$-2$ channels, \emph{i.e.}~$E^{v'-1}_\mathrm{avail}$ and
$E^{v'-2}_\mathrm{avail}$.

\begin{table}[!t]
 \begin{center}
  \caption{Energies (in cm$^{-1}$) in the vibrational predissociation of
   Ne$_2$Br$_2$($B,v'$) (see text for full details). \label{tab:energetics}}
  \begin{tabular}{crccrr}
   \hline\hline
   $v'$ & $\Delta\mathcal{E}^{B,v'-1}_\mathrm{Br_2}$ & $D^{\mathrm{NeBr_2}(B,v')}_0$ & $D^{B,v'}_0$ & $E^{v'-1}_\mathrm{avail}$ & $E^{v'-2}_\mathrm{avail}$ \\ \hline
   16   & 107.4                                      & 63.3                          & 137          & 33                        & 81                        \\
   17   & 103.8                                      & 63.1                          & 137          & 29                        & 74                        \\
   18   & 100.2                                      & 62.8                          & 137          & 26                        & 67                        \\
   19   & 96.6                                       & 62.6                          & 137          & 22                        & 60                        \\
   20   & 93.1                                       & 62.3                          & 136          & 19                        & 54                        \\
   21   & 89.5                                       & 62.0                          & 136          & 15                        & 47                        \\
   22   & 85.9                                       & 62.5                          & 135          & 13                        & 40                        \\
   23   & 82.3                                       & 62.4                          & 135          &  9                        & 33                        \\
   \hline\hline
  \end{tabular}
 \end{center}
\end{table}

Tests on basis-set convergence yield an estimate for the accuracy of our
$D^{B,v'}_0$ values of about 1~cm$^{-1}$ (any further improvement was considered
unnecessary for our purposes here).  These theoretical values are just slightly
different from the 141 cm$^{-1}$ estimate of Pio \emph{et al.}~\cite{jmpio:10},
and suggest an interaction energy for the Ne--Ne bond of about 10--12~cm$^{-1}$,
which is to be compared with the 17 cm$^{-1}$ of isolated Ne$_2$ used in
Ref.~\onlinecite{jmpio:10}.  Our predictions are, however, consistent with the
value of 12.81~cm$^{-1}$ in Ref.~\onlinecite{bmiguel:01} for the
effective Ne--Ne bond.  Moreover, our calculations predict an average Ne--Ne
interatomic distance at the minimum energy structure of about 3.2~\AA~(within
3.8\% of the value for Ne$_2$, see Table~\ref{tab:Morse}), also in good
agreement with the estimate in Ref.~\onlinecite{bmiguel:01}.  It is important to
note that all values in Table~\ref{tab:energetics}, except for
$D^{\mathrm{NeBr_2}(B,v')}_0$, are theoretical predictions based on the
$B$-state PES used in this work.  This is the reason for small discrepancies
with some experimental predictions \cite{jmpio:10}.  For instance, we estimate
the closing of the $\Delta v' = -1$ channel for the dissociation of the first Ne
atom to occur at $v' = 25$, as opposed to the experimental $v' = 23$.  Also,
complete dissociation via the $\Delta v' = -2$ channel would be energetically
accessible up to $v' = 27$ (instead of $v' = 25$).

\subsubsection{Vibrational branching ratios}
\label{sec3:Pv}
Experimental and QCT vibrational branching ratios after the dissociation of one
Ne atom, NeBr$_2$($v'-n$):NeBr$_2$($v'-1$), and two Ne atoms,
Br$_2$($v'-n$):Br$_2$($v'-2$), are given in Table~\ref{tab:Pv}.
\begin{table*}[!t]
 \begin{center}
  \caption{Experimental \cite{jmpio:10} and theoretical (this work) branching
   ratios in the vibrational predissociation of Ne$_2$Br$_2$($B,v'$).
   \label{tab:Pv}}
  \begin{tabular}{clcccccccc}
   \hline\hline
        &            & \multicolumn{3}{c}{NeBr$_2$($v'-n$):NeBr$_2$($v'-1$)} && \multicolumn{4}{c}{Br$_2$($v'-n$):Br$_2$($v'-2$)} \\ \cline{3-5}\cline{7-10}
   $v'$ &            & ($v'-2$)  & ($v'-3$)  & ($v'-4$)                      && ($v'-1$) & ($v'-3$)  & ($v'-4$)  & ($v'-5$)       \\ \hline
   16   & Experiment &           &           &                               &&          & 0.16      &           &                \\
        & QCT: SB/GW & 0.04/0.27 &           &                               && 0.98/0.0 &           &           &                \\ \hline
   17   & Experiment &           &           &                               &&          & 0.31      & 0.05      &                \\
        & QCT: SB/GW & 0.08/0.29 &           &                               && 0.48/0.0 &           &           &                \\ \hline
   18   & Experiment &           &           &                               &&          & 0.33      & 0.08      &                \\
        & QCT: SB/GW & 0.16/0.35 &           &                               && 0.20/0.0 & 0.01/0.0  &           &                \\ \hline
   19   & Experiment &           &           &                               &&          & 0.45      & 0.09      &                \\
        & QCT: SB/GW & 0.28/0.45 &           &                               && 0.07/0.0 & 0.02/0.01 &           &                \\ \hline
   20   & Experiment &           &           &                               &&          & 0.65      & 0.13      & 0.03           \\
        & QCT: SB/GW & 0.47/0.56 & 0.01/0.01 &                               && 0.02/0.0 & 0.05/0.03 &           &                \\ \hline
   21   & Experiment &           &           &                               &&          & 0.97      & 0.29      &                \\
        & QCT: SB/GW & 0.71/0.73 & 0.02/0.02 &                               && 0.01/0.0 & 0.10/0.06 &           &                \\ \hline
   22   & Experiment &           &           &                               &&          & 1.58      & 0.51      & 0.17           \\
        & QCT: SB/GW & 1.04/0.92 & 0.06/0.04 &                               && 0.01/0.0 & 0.18/0.10 & 0.01/0.0  &                \\ \hline
   23   & Experiment &           &           &                               &&          & 3.28      & 1.37      & 0.49           \\
        & QCT: SB/GW & 1.52/1.22 & 0.15/0.09 & 0.01/0.0                      &&          & 0.29/0.15 & 0.02/0.01 &                \\
   \hline\hline
  \end{tabular}
 \end{center}
\end{table*}
It is readily apparent that QCT fails to capture the physics behind the
vibrational distributions in the VP process, irrespectively of the statistical
procedure (SB or GW) employed.  This is a general effect of the large mismatch
in the strengths of the different DOF involved in the VP of vdW clusters, at least for
moderate vibrational excitations of the chemically-bounded diatom. The
same behavior may therefore be expected in similar systems.  In the absence of
resonances, such a mismatch causes the energy transfer from the vibrational to
the vdW dissociation modes to be very inefficient.  This is remarkably different
from the unimolecular dissociation of more conventional, \emph{i.e.}\
chemically-bounded, molecules, where statistical arguments are usually
applicable at least for the DOF directly involved in the fragmentation.  In the
classical description of the VP of Ne$_2$Br$_2$, the Br$_2$ vibrational energy
gradually  `flows' into the vdW modes and eventually becomes large
enough for dissociation to occur.  A very limited number of the accessible final
vibrational states is hence populated, the distributions being highly
non-statistical (Fig.~\ref{fig:tdissvf_23dof_v23}).  In addition, due to the
very low density of vibrational states, the classical picture dramatically
differs from the quantum-mechanical description and hence the experimental
observations.  This is particularly clear from SB product branching ratios.  In
this case, the loss of 1.5 `quanta' provides enough energy to eject both Ne
atoms up to $v' = 22$ (Table~\ref{tab:energetics}), which is reflected in
nonphysical trajectories dissociating via the $\Delta v' = -1$ channel
(Table~\ref{tab:Pv}).  The main advantage of GW over SB results in this case is
to populate the qualitatively correct vibrational channels.  However, as
discussed above, the classical description is intrinsically inadequate and GW
predicts a larger than observed propensity for $\Delta v' = -1$ and $-2$
channels for the respective dissociation of the first and second Ne atoms.
Also, the increasing importance of highly state-specific IVR (sparse regime)
predicted in the experiment for states above $v' = 19$ cannot be adequately
reproduced within QCT.  Some additional discrepancies, \emph{e.g.}\ in channels
closings, are due to the PES employed and have been already discussed at the end
of Sec.~\ref{sec2:pdistros}.

\subsubsection{Rotational distributions}
\label{sec3:Pj}
\begin{figure}[!t]
 \begin{center}
  \includegraphics[width=85mm]{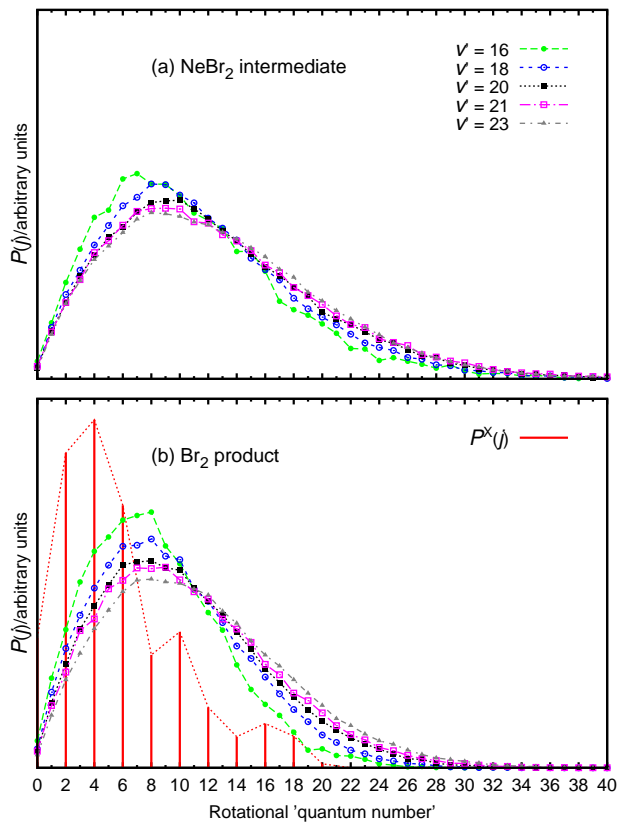}
  \caption{(Color online) Rotational distributions in the vibrational
   predissociation of Ne$_2$Br$_2$($B$), for: (a) NeBr$_2$ intermediate
   complex; and (b) Br$_2$ diatomic product.  $P^X(j)$ is the quantum
   distribution used to generate the quasi-classical initial conditions.
   \label{fig:Pjall}}
 \end{center}
\end{figure}

Together with vibrational distributions, product rotational state distributions
provide a more detailed picture of the VP process than dissociation lifetimes.
In particular, they contain important information on the anisotropy of the PES,
the vibrational level spacings and the effects of IVR and resonances on the
fragmentation dynamics \cite{rschinke:book93}.  Rotational distributions
corresponding to various initial $v'$ levels for the NeBr$_2$ intermediate
triatomic complex and the Br$_2$ diatomic product are depicted in
panels a and b of Fig.~\ref{fig:Pjall} respectively.  Only the GW results are
shown because even if just a few vibrational states are populated, the SB and GW
curves are only slightly different.  This is not particularly surprising;
although the available energy more than halves in the range of $v'$ considered
(Table~\ref{tab:energetics}), the tail of QCT distributions rapidly tend to zero
before this effect becomes important.

The rotational state of the NeBr$_2$ intermediate complex was calculated from
$\boldsymbol{j} + \boldsymbol{l}_{1(2)} = -\boldsymbol{l}_{2(1)}$, which is
valid only for $\boldsymbol{J} = \boldsymbol{0}$.  As seen in panel (a), the
dissociation of the first Ne atom leaves the resultant triatomic complex in a
highly-excited rotational state.  The corresponding distribution extends over
more than 30 rotational levels and peaks around 8--10. Such rotational
excitation, together with the additional excitation of the vdW vibrations, are
expected to be the cause for the lifetime of the intermediate molecule being
larger than that of the directly excited triatomic complex.  This effect has
been observed in the experiment \cite{jmpio:10} and is reproduced by our QCT
calculations, as seen by comparing the corresponding lifetimes in
Table~\ref{tab:lifetimes} with the QCT predictions in Table~1 of
Ref.~\onlinecite{mlglez-mtnez:06a}.

The GW rotational distributions for the Br$_2$ product, shown in panel (b),
extend over the full range of accessible states determined by the maximum
available energy in Table~\ref{tab:energetics}.  For the sake of comparison, the
initial distribution $P^X(j)$ has also been included in this panel.  In the
particular case of $\boldsymbol{J} = \mathbf{0}$, symmetry considerations for
the ground state are only responsible for even $j$ values contributing in
$P^X(j)$ \cite{pvillarreal:94}.  In general, the calculated rotational
distributions are considerably hotter for the products, with a maximum at about
$j^\mathrm{f} = 6$--8 almost independent of $v'$, than for the initial states.
The initial and final distributions have, nevertheless, similar shapes.
Although experimental rotational distributions were not reported in the
experiment \cite{jmpio:10}, a few points can be made with respect to QCT
predictions: (a) the Boltzmann-like shape seems to be consistent with
experimental measurements for the directly excited NeBr$_2$ molecule
\cite{mnejad-sattari:97}, and previous QCT calculations \cite{mlglez-mtnez:06a,
mlglez-mtnez:08}; and (b) the Br$_2$ product is more rotationally excited than
in the photo-dissociation of NeBr$_2$\cite{mnejad-sattari:97,mlglez-mtnez:06a,
mlglez-mtnez:08}, for which the distributions peak about $j^\mathrm{f} = 4$--6.
The extent of the QCT distributions may be expected to differ from future
experimental measurements due to the PES employed (see discussion at the end of
Sec.~\ref{sec2:pdistros}), particularly close to a channel closing.

\section{Summary and conclusions}
\label{sec:summary+conclusions}
We have studied the vibrational predissociation (VP) of the Ne$_2$Br$_2$($B$,
$v'=$~16--23) van der Waals (vdW) cluster using the quasi-classical trajectory
(QCT) and the (Cartesian) coupled coherent states (CCS) methods.

A sequential mechanism was used to fit the dynamical evolution of the different
complexes involved.  Both QCT and CCS are shown to provide very good estimates
for the different dissociation lifetimes reported in the experiment
\cite{jmpio:10} and previous molecular dynamics with quantum transitions (MDQT)
simulations \cite{bmiguel:01}.  QCT predictions are, however, obtained at a much
lower computational expense.  Various sources of non-exponential behavior have
been identified and their implications extensively discussed.  In particular,
the initial shape of the QCT curves at short times arises from the
`non-democratic' selection of classical initial conditions and the classical
description of the process.  Meanwhile, the quantum CCS simulations give curves
whose initial shape is in much better agreement with the experimental
observations \cite{jmpio:10}.  However, the basis functions used to describe the
wave function follow dissociative trajectories.  The high dimensionality of the
phase space therefore results in the basis functions decoupling and thus a
semi-classical description of the cluster.  This change is as least partially
responsible for the departure at longer times of the curves from their initial
exponential shape.  An intrinsic multi-step dissociation mechanism, as observed
in the experiment \cite{jmpio:10} and predicted by MDQT calculations
\cite{bmiguel:01}, may also be responsible for such non-exponential behavior at
long times.  Additional calculations are needed in order to clarify this issue.
The time scale over which the CCS gives a good description was increased
significantly to around 15~ps by re-expanding the dissociated basis functions.
Despite this improvement, the QCT simulations give curves that agree much more
closely with the experimental results at longer times.  The behavior of the QCT
curves at longer times is, however, affected by low-dimensional tori in
classical phase space.  These tori are mainly in the form of quasi-periodic
trajectories and are partially due to the weakness of vdW interactions.  In both
QCT and CCS calculations, the percentage of trajectories which correspond to the
concerted mechanism is practically negligible, below 0.1\% for QCT and about 2\%
for CCCS.  In this regard, it is important to note that trajectory-based
approaches like CCS may become more effective than `standard' quantum methods,
mainly because they allow analyzing the different mechanisms by simply
inspecting the trajectories.

As in our previous work with QCT on triatomic vdW molecules
\cite{mlglez-mtnez:08}, we found that application of the Gaussian weighting (GW)
procedure yields survival probability curves, and consequently lifetimes, which
are not significantly different from those calculated using the standard binning
(SB) procedure.  In addition, the capabilities of QCT in the description of the
fragmentation kinetics was analyzed in detail by using reduced-dimensionality
models of the complexes and concepts from phase-space transport theory.

We have reported QCT ro-vibrational product state distributions for the
intermediate and final states of the VP process, and compared the vibrational
distributions with the experimental results of Pio \emph{et al.}\
\cite{jmpio:10}.  As in previous studies,
\emph{e.g.}~Ref.~\onlinecite{mlglez-mtnez:08}, the SB was found to populate
nonphysical dissociation channels ($\Delta v' = - 1$ in this case).  This is due
to energetic issues and can be easily solved by using the GW method.  The
latter, however, predicts a larger than observed propensity for dissociation of
the first (second) Ne atom via the $\Delta v' = -1$ (-2) channel.  We argued
that this is a general problem in the classical description of the VP of vdW
clusters, which may be attributed to the weakness of vdW interactions.  Due to
the latter, the characteristic frequencies for the diatomic subunit are usually
one or more orders of magnitude larger than those of the bending mode within the
vdW well.  The energy transfer leading to dissociation is inefficient and slow.
To complicate matters further, as energy flows from the vibrational to the
transitional modes, the vdW bending modes evolve from bounded, through hindered,
to free type of motion and the role of many non-linear resonances become
increasingly important \cite{aagranovsky:98}.  This fact additionally worsens
the quality of any classical description of the VP process, since quantum
state-selectivity cannot be adequately described.  As expected, the
quantum-classical discrepancies will be more pronounced for relatively low $v'$,
given the very low density of vibrational states.

Dissociation of the first Ne atom leaves the NeBr$_2$ in a highly-excited
rotational state.  This rotational excitation, and the additional excitation of
the vdW vibration, are responsible for an increase dissociation lifetime of the
intermediate complex, as compared to the lifetime of the directly excited
NeBr$_2$.  This effect was observed in the experiment \cite{jmpio:10} and is
correctly predicted by our calculations.  Product Br$_2$ rotational
distributions are found to be significantly hotter in the products, extending
over the full range of available energies.

\begin{acknowledgments}
WAG, MLGM and JRS wish to thank Prof.~Gerardo Delgado-Barrio, from the Consejo
Superior de Investigaciones Cient\'{\i}ficas, Espa\~na, for his contribution in
the calculation of quasi-classical initial distributions.  These authors wish
also to acknowledge the support from the PNCB/2/9 project of the Departamento de
F\'{\i}sica General in the Instituto Superior de Tecnolog\'{\i}as y Ciencias
Aplicadas, Cuba.  SKR would like to thank the EPSRC for funding through grant
number EP/E009824/1 and more recently the University of Leeds for a Visiting
Research Fellowship.  SKR would also like to acknowledge the use of the UK
National Grid Service (NGS) and the University of Leeds central HPC system in
performing the CCCS simulations.  The collaboration between the Cuban and
British groups was possible thanks to an International Joint Project grant from
the Royal Society.
\end{acknowledgments}

%

\end{document}